\documentclass[useAMS,usenatbib]{mn2e}
\usepackage{graphicx}
\usepackage{amsmath}
\usepackage{color}
\usepackage{amssymb}
\usepackage[table]{xcolor}

\title[Machine-z]{Machine-z: Rapid Machine Learned Redshift Indicator for $Swift$ Gamma-ray Bursts}
\author[T. N. Ukwatta, P. R. Wo\'zniak and N. Gehrels]{
T. N. Ukwatta$^{1}$\thanks{E-mail: tilan@lanl.gov; tilan.ukwatta@gmail.com}, P. R. Wo\'zniak$^{2}$ and N. Gehrels$^{3}$\\
$^{1}$Director's Postdoctoral Fellow, Space and Remote Sensing (ISR-2), Los Alamos National Laboratory, Los Alamos, NM 87544, USA.\\
$^{2}$Space and Remote Sensing (ISR-2), Los Alamos National Laboratory, Los Alamos, NM 87544, USA.\\
$^{3}$Astroparticle Physics Division, NASA/Goddard Space Flight Center, Greenbelt, MD 20771, USA.}

\begin{document}

%\date{Accepted 1988 December 15. Received 1988 December 14; in original form 1988 October 11}

%\pagerange{\pageref{firstpage}--\pageref{lastpage}} \pubyear{2002}

\maketitle

\label{firstpage}

\begin{abstract}
Studies of high-redshift gamma-ray bursts (GRBs) provide important information about
the early Universe such as the rates of stellar collapsars and mergers,
the metallicity content, constraints on the re-ionization period, and probes of
the Hubble expansion. Rapid selection of high-z candidates from GRB samples
reported in real time by dedicated space missions such as $Swift$ is the key
to identifying the most distant bursts before the optical afterglow becomes
too dim to warrant a good spectrum. Here we introduce ``machine-z'',
a redshift prediction algorithm and a ``high-z'' classifier for $Swift$ GRBs based on machine learning.
Our method relies exclusively on canonical data commonly available within
the first few hours after the GRB trigger. Using a sample of 284 bursts with measured
redshifts, we trained a randomized ensemble of decision trees (random forest)
to perform both regression and classification. Cross-validated performance studies
show that the correlation coefficient between machine-z predictions and the true redshift
is nearly 0.6. At the same time our high-z classifier can achieve 80\% recall of true
high-redshift bursts, while incurring a false positive rate of 20\%. With
40\% false positive rate the classifier can achieve $\sim$100\% recall.
The most reliable selection of high-redshift GRBs is obtained
by combining predictions from both the high-z classifier and the machine-z regressor.
\end{abstract}

\begin{keywords}
gamma-ray bursts, redshift
\end{keywords}

%\tableofcontents

\section{Introduction}

Gamma-ray bursts (GRBs) are often characterized as the most energetic electromagnetic
explosions since the beginning of the Universe.
Their optical afterglows are in principle detectable out to
redshift $z > 10$ \citep{lamb2000,Mesler2014}. Therefore, studies of distant GRBs can probe
the physics of the early Universe including the re-ionization, the evolution
of star formation, and the process of metal enrichment \citep{lamb2000,Totani2006,Kawai2006}.
High-redshift GRBs can be used to pinpoint and characterize the faint galaxies
that supplied most of the re-ionization photons and to constrain the re-ionization
redshift~\citep{Ioka2003, Totani2006, Wang2012}. Multi-wavelength studies of distant GRB afterglows
can provide new information about the metal and dust content of these objects
~\citep{Frail2006, Cusumano2006, Mesler2014}. This in turn offers a unique method to
understand the metal enrichment history of sources during the re-ionization epoch.
While most studies of the re-ionization epoch are based on observations
of quasars, high-z GRBs have a number of significant advantages over quasars
due to their unique characteristics. GRB afterglows are exceptionally bright
and provide plenty of photons for sensitive spectroscopy. They have simple, easy to model
power-law spectra dominated by the continuum emission that are well suited for detecting
absorption signatures of the intergalactic medium. Additionally, the neighborhoods
of GRB progenitors are relatively ``clean'' compared to quasars,
which are often contaminated by continuous ejection of material
from the central engine.

The $Swift$ Gamma-Ray Burst Mission \citep{Gehrels2004} has proven to
be effective in detecting very high-redshift GRBs. The most distant spectroscopically
confirmed GRB on record is GRB 090423 with $z = 8.2$
\citep{Tanvir2009, Salvaterra2009}\footnote{GRB 120923A may have slightly higher redshift.
A preliminary analysis of the photometric data indicates $z\sim8.5$~\citep{Tanvir2013}.}.
The highest photometrically measured burst is GRB 090429B with a redshift
of 9.4~\citep{Cucchiara2011}. There is now a handful of spectroscopically confirmed
GRBs with $z > 5$. The main challenge in this work is to reliably identify high-redshift
bursts suitable for detailed spectroscopic follow-up before the optical emission fades away.
Decisions to use precious observing time on large telescopes must be made within
the first hours or even minutes after the burst based on limited information.
Previous attempts to screen high-z GRBs using promptly available high-energy
data~\citep{Campana2007,Salvaterra2007,Koen2009,Koen2010,Ukwatta2008,Ukwatta2009,Morgan2012},
while showing some promise, lacked the accuracy necessary to facilitate a reliable
follow-up program. As a result, they were never widely adopted by observers.

The main difficulty lies in extracting numerous weak correlations from readily available
high-dimensional data and efficiently combining the information they contain.
A modern approach based on machine learning is ideal for this purpose. Starting from
a catalog of GRBs with known redshifts we can use supervised learning to ``train''
algorithms that effectively encode the relationship between input data and output labels.
Classification algorithms deal with predicting discrete labels (in this case high-z versus low-z),
while regression algorithms predict continuous labels (here the redshift value).
Both types of models are supported by the random forest algorithm that in recent years
has emerged as one of the best performing machine learning tools in observational astrophysics.

In this paper we present a rapid machine learned redshift estimator called {\it machine-z}
and a {\it high-z} classifier for GRBs detected by $Swift$. Both {\it machine-z} and {\it high-z}
are developed independently and each tool may be used to reinforce conclusions from the other.
In Section~\ref{methodology} we describe the input data and the method. Our high-z classifier is
developed in Section~\ref{highz_classifier} and the machine-z indicator is
developed in Section~\ref{machine_z}. In Section~\ref{discussion} we compare
our algorithms and results with previous work and evaluate the performance of
our new tools using a sample of recently detected bursts that are not included
in the training catalog. In Section~\ref{summary} we summarize the results.

\section{Methodology} \label{methodology}

\subsection{GRB Sample} \label{grb_sample}

Our sample consists of 284 $Swift$ GRBs with spectroscopic redshift
measurements\footnote{The measurements are taken from the $Swift$
online catalog at \texttt{http://swift.gsfc.nasa.gov/archive/grb\_table/}.}.
The $Swift$ mission payload consists of three major instruments: the Burst Alert Telescope (BAT),
the X-ray Telescope (XRT) and the UV Optical Telescope (UVOT)~\citep{Gehrels2004}.
BAT is a soft gamma-ray wide field instrument sensitive to photons in the energy
range 15 keV to 350 keV and it is the GRB discovery instrument. Once BAT discovered a GRB and
determined its sky position, the $Swift$ satellite slews to the location of the burst
so that the narrow field instruments XRT and UVOT can quickly start observing the afterglow.
In order to provide a rapid {\it machine-z} redshift and {\it high-z} classification,
we limited this study to readily available measurements from all three $Swift$ instruments.
These measurements were adopted as numerical features for classification and regression,
and are listed in Table~\ref{features}.

In total we considered 25 features. The features generally capture the timing and
spectral properties of the prompt and afterglow emission from the burst.
Some measurements are encoded as two separate features to ensure that all available
information is included. For example, the prompt emission recorded by BAT is fitted using
either a power law (PL) model or a cutoff power law (CPL) model, depending on which one
provides a better fit. The power-law index is an important feature, but the existence
of a high-energy cutoff provides additional information and is included as a separate
feature {\tt FitType} that encodes the best fit model (0 for PL and 1 for CPL).
Similarly, UVOT magnitudes are reported as either detections or upper limits.
A numerical flag similar to {\tt FitType} is introduced to include this information
for each photometric band of UVOT.

\begin{table*}
\centering
 \caption{Standard features used for training classification and regression algorithms.
The features are derived from measurements by all three $Swift$ instruments: BAT, XRT and UVOT.}
 \label{features}
 \begin{tabular}{@{}rlcc}
  \hline \hline
  Item & Feature & Units & Instrument \\
  \hline \hline
 1 &  T90	& sec & BAT \\
 2 &  Fluence (15-150 keV)	& $10^{-7} \rm erg/cm^2$ & BAT \\
 3 &  1-sec Peak Photon Flux (15-150 keV) & $\rm ph/cm^2/sec$ & BAT \\
 4 &  Photon Index	& None & BAT \\
 5 &  Fit Type (CPL - Cutoff Power Law, PL - Power Law) & CPL=1, PL=0 & BAT \\
 6 &  Early Flux (0.3-10 keV) & $10^{-11} \rm erg/cm^2/s$ & XRT \\
 7 &  11 Hour Flux (0.3-10 keV) & $10^{-11} \rm erg/cm^2/s$ & XRT \\
 8 &  24 Hour Flux (0.3-10 keV) & $10^{-11} \rm erg/cm^2/s$ & XRT \\
 9 &  Initial Temporal Index & None  & XRT \\
10 &  Spectral Index ($\Gamma$) & None  & XRT \\
11 &  Column Density (NH) & $10^{21} \rm cm^{-2}$ & XRT \\
12 &  V Mag/Limit Value & Magnitudes & UVOT \\
13 &  V Value Type	& Mag=1, Limit=0 & UVOT \\
14 &  B Mag/Limit Value & Magnitudes & UVOT \\
15 &  B Value Type	& Mag=1, Limit=0 & UVOT \\
16 &  U Mag/Limit Value & Magnitudes & UVOT \\
17 &  U Value Type	& Mag=1, Limit=0 & UVOT \\
18 &  UVW1 Mag/Limit Value & Magnitudes & UVOT \\
19 &  UVW1 Value Type	& Mag=1, Limit=0 & UVOT \\
20 &  UVM2 Mag/Limit Value & Magnitudes & UVOT \\
21 &  UVM2 Value Type	& Mag=1, Limit=0 & UVOT \\
22 &  UVW2 Mag/Limit Value & Magnitudes & UVOT \\
23 &  UVW2 Value Type	& Mag=1, Limit=0 & UVOT \\
24 &  White Mag/Limit Value & Magnitudes & UVOT \\
25 &  White Value Type	& Mag=1, Limit=0 & UVOT \\
  \hline
 \end{tabular}
\end{table*}

\begin{figure}
\includegraphics[width=84mm]{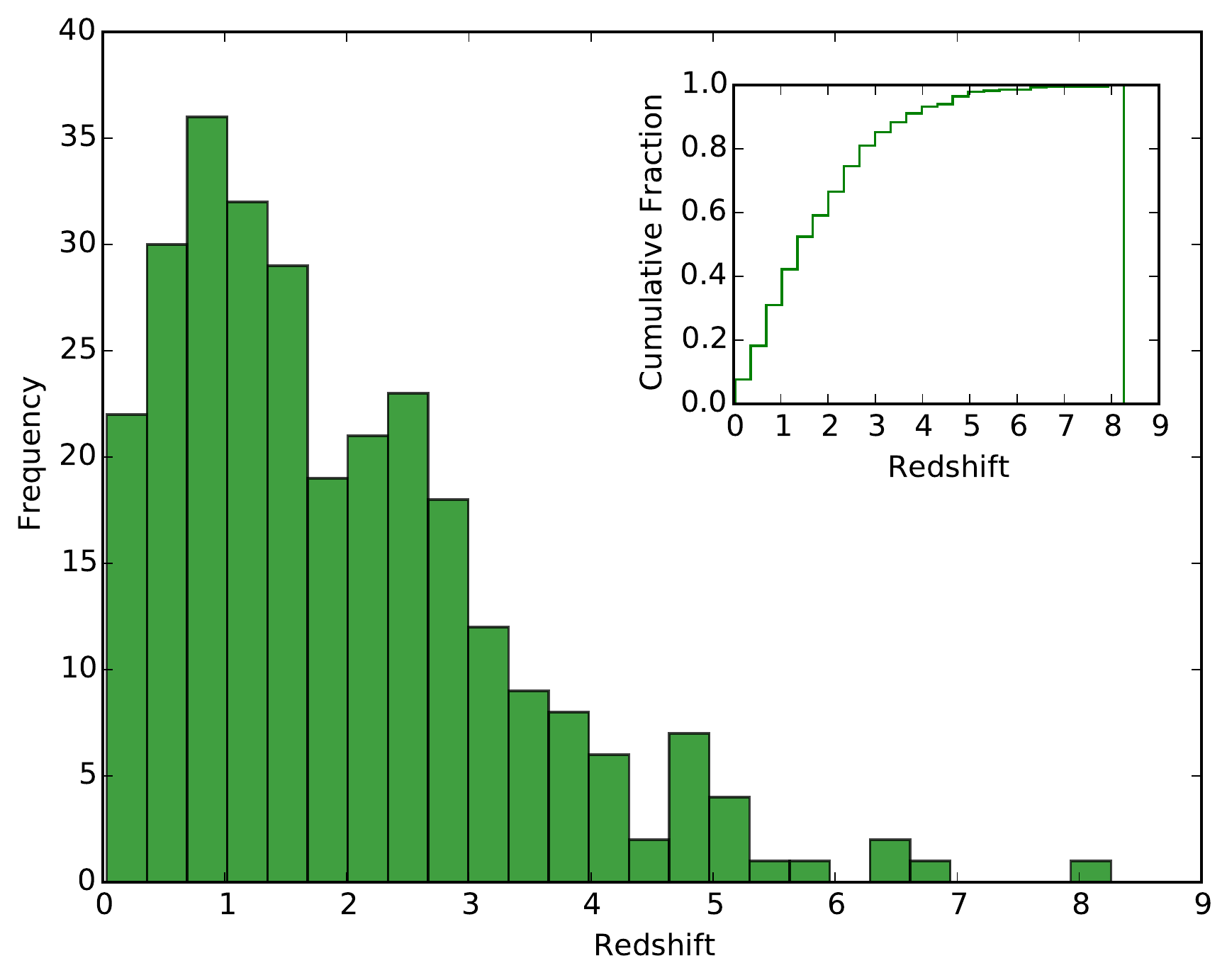}
\caption{Redshift distribution for a sample of 284 $Swift$ bursts.}\label{zhisto}
\end{figure}

The sample includes GRBs discovered between 2005 and the end of 2014.
Fig.~\ref{zhisto} shows the redshfit distribution of all bursts in the sample.
The lowest redshift value is 0.033 and the highest is 8.26.
We adopt $z = 4$ as the threshold between low-redshift and high-redshift bursts.
Out of 284 GRBs in the sample 25 or $\sim9$\% are high-redshift according to this definition.

\subsection{Machine Learning Algorithm}

\subsubsection{Random Forests}

The random forest (RF) algorithm~\citep{Breiman2001} has been shown to provide superior performance 
on many classification problems~\citep{Caruana2006, duBuisson2015, DIsanto2016}. Over the past few years the method found several
interesting applications in observational astrophysics including selection of explosive transients 
in imaging data~\citep{Wright2015, Goldstein2015}, classification of X-ray sources~\citep{Farrell2015}, 
and redshift prediction~\citep{Carliles2010, Morgan2012}. RF has the ability to select useful features,
relatively immune to data over-fitting, can handle nonlinear relationships, 
and provide probabilistic outputs (see \cite{Morgan2012} and references therein).
Given input training data the algorithm creates a large number of decorrelated binary decision trees.
Each tree in the forest is grown by splitting the portion of the training data associated
with a particular parent node between two child nodes according to the value of one
or more features. Splits are chosen to maximize node purity (classification) or minimize
variance (regression). The process starts from the root node that holds the entire data set
and continues until the size of the leaf node falls below a specified threshold.
Randomness enters in two distinct ways.
A new bootstrap sub-sample is drawn from the original training sample to train the next tree.
Then a random subset of all available features is selected to optimize each split. The size
of the random subset is specified by the user.

Prediction for a single tree is accomplished by propagating a previously unseen feature
vector starting from the root and until a leaf node is reached. The predicted label
is typically a majority class of the leaf node (classification) or a numerical average
(regression). The results from all trees in the forest are then combined to compute
the posterior probability of possible outcomes and the final prediction.
Typically the majority vote is adopted for classification and the mean for regression.
All results described in this paper were obtained using the python implementation
of RF distributed with the scikit-learn 
package\footnote{\texttt{http://scikit-learn.org}}~\citep{Pedregosa2012}.

\subsubsection{Missing Features}

Missing features are common in real world data and our GRB sample is
no exception. A popular approach to handle missing input values is by
imputation i.e. assigning values estimated from the distribution of
all remaining instances. The method works well unless missing
values carry a special meaning in a given particular problem domain.
In our input GRB catalog some features could not be extracted due
to low signal to noise ratio of the original data or a non-detection
in a particular photometric band. Both occurrences are actually expected to
be correlated with redshift (distance) and are therefore examples
of informative missing features. Blue filter ``dropouts'' are especially
interesting because they are often indicative of high redshift.
In order to preserve all available information about the redshift
we assigned all missing features to $-1000$. The effect on performance is
negligible as long as the plug-in value is well outside
the normal range for all features.
Decision trees are generally very good in utilizing such special values
if the missing features are in fact informative or marginalizing them out if they are not.

\subsubsection{Data Imbalance}
\label{imbalance}

Another common problem when searching for rare objects of interest is
highly uneven distribution of training data between classes.
The redshift distribution in Fig.~\ref{zhisto} represents competition
between survey volume growing rapidly with distance and decreasing
efficiency of detecting more distant bursts. The input catalog for our study
is quite unbalanced with fewer than 10\% of GRBs at $z > 4$.
A low fraction of high-redshift
bursts in the training sample will typically result in a tendency to classify
all bursts as low-redshift as the algorithm attempts to minimize the overall
error rate. This imbalance will result in poor performance of the classifier
on new data. One possible solution is to assign higher weights to high-z bursts
during training. However, the price is often additional complexity
and a tendency for overfitting. It is not clear at this point what is the optimal
way to introduce weights in random forest and almost certainly the answer depends on
the problem at hand. A preliminary investigation of the redshift bias
discussed in Section~\ref{machinez_cal} indicates that the underlying cause
is imbalanced training data combined with noisy features in a significant fraction
of the sample. A detailed treatment of these intricate effects warrants a separate
investigation and will be presented elsewhere.

%/home/tilan/Desktop/Dropbox/grb/high_z_screening/roc_curve_analysis/cv_analysis_k_fold_roc_v2.py
\begin{figure*}
\includegraphics[width=0.8\textwidth]{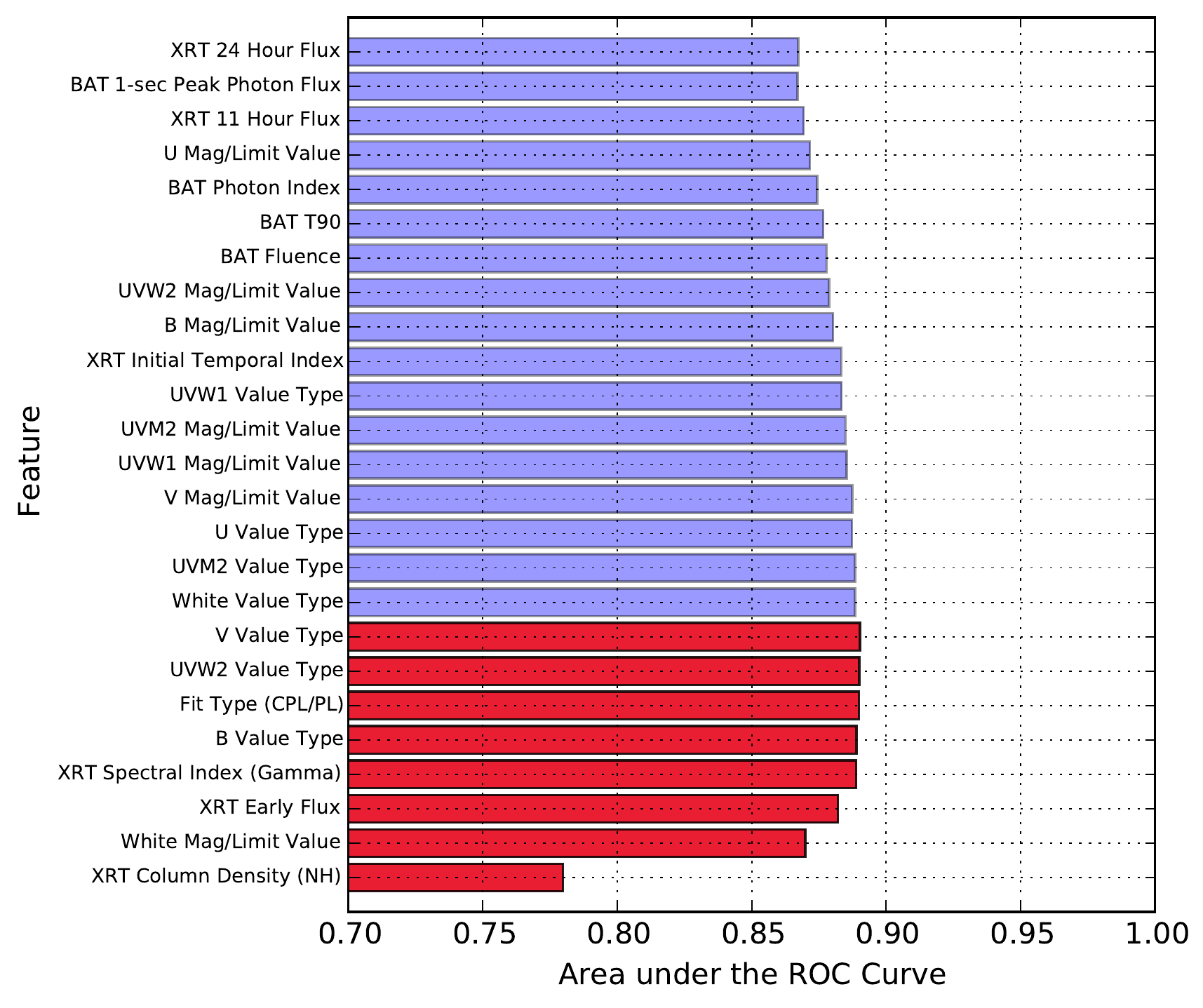}
\caption{Relative importance of {\it high-z} classification features.
The area under the ROC curve is shown as a function of the next best feature
starting from the single best feature at the bottom of the plot.
Features selected for the final {\it high-z} classifier are shown in red.
}\label{feature_selection_classifier}
\end{figure*}

\section{High-z Classification} \label{highz_classifier}

\subsection{Receiver Operating Characteristic (ROC) Curve}
\label{roc_curve}

The receiver operating characteristic (ROC) curve is a convenient way to
track performance and compare classifiers and/or feature sets. We compute this curve
using randomized 10-fold cross-validation (train on 90\% of the sample and test on 10\%).
A single point on the ROC curve corresponds to an average of 100 independent
cross-validation runs for a fixed threshold applied to the probability of the burst
having a high redshift. The curve is traced by varying the threshold between 0 and 1
(see Fig.~\ref{roc_curve} for an example). A perfect classifier
has zero false positive rate and 100\% efficiency (upper left corner of the diagram).
Fast rising ROC curves are generally better
than a slow rising ones. The area under the curve can be used as a rough
measure of classification performance. An ideal classifier has the area
of 1, while completely random selection on average yields half of the total area
of the diagram.

\subsection{Tuning the Classifier}
\label{tune_classifier}

RF classifiers take several parameters that can be tuned to improve performance.
Among those the most important are the number of trees in the forest ($ntrees$),
the minimum size of the leaf node ($nodesize$), and the number of randomly selected
features that will be used to optimize node splits ($m$). In order to approximately
optimize the {\it high-z} classifier we performed a simple parameter search over a grid
given in Table~\ref{parameters} and selected a set of parameters with the largest
area under the ROC curve. This results in a forest of 300 trees with at least 12 training
samples per node and $m = 25$ random features per split that delivers an ROC curve
with an area of 0.87. The fact that $m = 25$ is preferred means that the best
results are obtained with a large degree of randomness injected during construction
of individual trees (all available features are randomized). The area under the ROC curve
is somewhat insensitive to the exact combination of parameters.

\begin{table}
\centering
 \caption{Parameter grid used to approximately optimize algorithm learning.}
 \label{parameters}
 \begin{tabular}{@{}ll}
  \hline \hline
  Parameter & Values \\
  \hline \hline
  $ntrees$ & 50, 100, 200, 300, 400, 500 \\
  $nodesize$ & 1, 2, 5, 8, 10, 12, 15, 18, 20 \\
  $m$ & 2, 5, 8, 10, 12, 15, 18, 20, 22, 25 \\
  \hline
 \end{tabular}
\end{table}

\subsection{Classification Feature Importance}
\label{feature_imp_classifier}

While RF is relatively immune to correlated and uninformative features,
it is still beneficial to investigate the relative importance of input features
on performance. We start with a pool of available features that initially
contains all features in Table~\ref{features}. The first most informative
feature is selected to maximize classification performance (area under ROC curve)
using only one feature at a time from the pool of $N$ available features.
The next best feature is selected after looping over $N-1$ features remaining in the pool
and maximizing classification performance using two features.
The process continues until the pool of available features is empty.
We use parameter values from Section~\ref{tune_classifier},
i.e. $ntrees = 300$ and $nodesize = 12$,
except $m$ which cannot be larger than the number of features selected
for a given iteration.
The relative importance of all 25 classification features
is shown in Figure~\ref{feature_selection_classifier}.
As more features are included in training, the area under the ROC curve increases rapidly
with a maximum value of 0.89 around 8-th feature followed by a gradual decrease.
It is interesting to note that the 8 best features selected in this way
include information from all three $Swift$ instruments. The absence of the total burst
duration (BAT T90) on this list is somewhat surprising and may be attributed to a large
intrinsic spread of burst durations that dominates the influence of time dilation
on this time scale.

The ROC curve of the final {\it high-z} classifier trained using the 8 best
features is shown on Fig.~\ref{roc_curve}. The curve shows
a steep rise and reaches 100\% recall at about
40\% false posive rate. We can reduce the false positive
rate by half by changing the probability threshold and accepting 80\% recall.

%/home/tilan/Desktop/Dropbox/grb/high_z_screening/roc_curve_analysis/combine_roc_curves.py
\begin{figure}
\includegraphics[width=84mm]{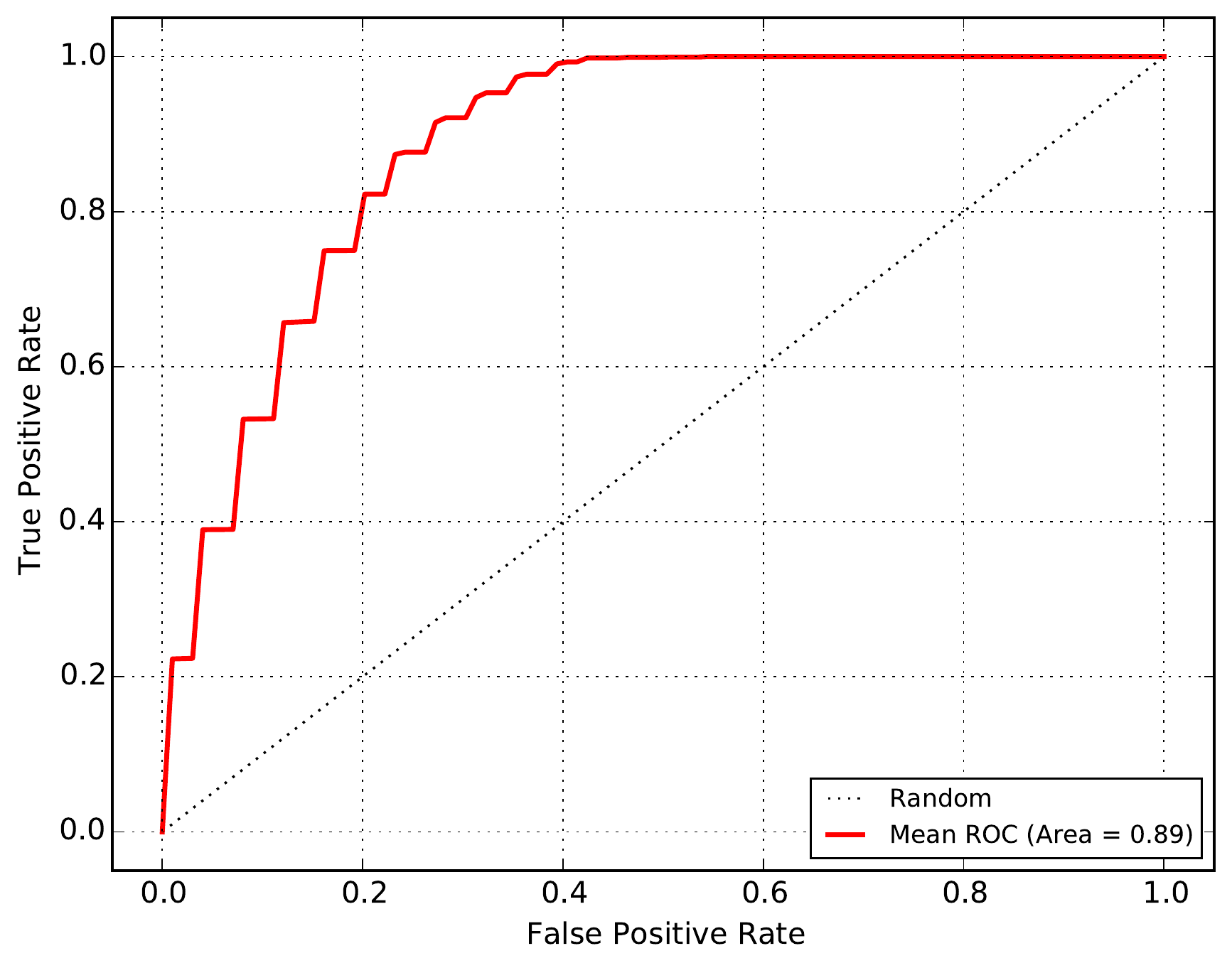}
\caption{ROC curve for {\it high-z} classifier using the best 8 features identified in
Fig.~\ref{feature_selection_classifier}.}\label{roc_curve}
\end{figure}

\subsection{Machine Learned Scoring}

The ROC curve is not the only way to measure the performance of selecting high-redshift GRBs.
Morgan et al. (2012) introduced a scoring method suitable
for scheduling follow-up observations constrained by limited resources (telescope time).
The idea is to compare {\it high-z} probabilities of all bursts in the training sample
to that of the new event under consideration. The new event is assigned a rank $n$,
with $n-1$ previously seen events having a higher probability of being high-z
compared to the new event. The learned follow-up rank of the new event is $Q=n/(N+1)$.
This formulation supports ``go or no go'' decisions for newly detected bursts,
given the fraction $F$ of all GRBs that can be followed up with the currently
available resources. If $Q < F$ the event should be observed. Otherwise it makes more sense
to wait for a better candidate. The algorithm automatically adapts to changes
in the redshift distribution of the input GRB sample and the amount of telescope time
that can be devoted to follow-up.

\begin{figure}
\includegraphics[width=84mm]{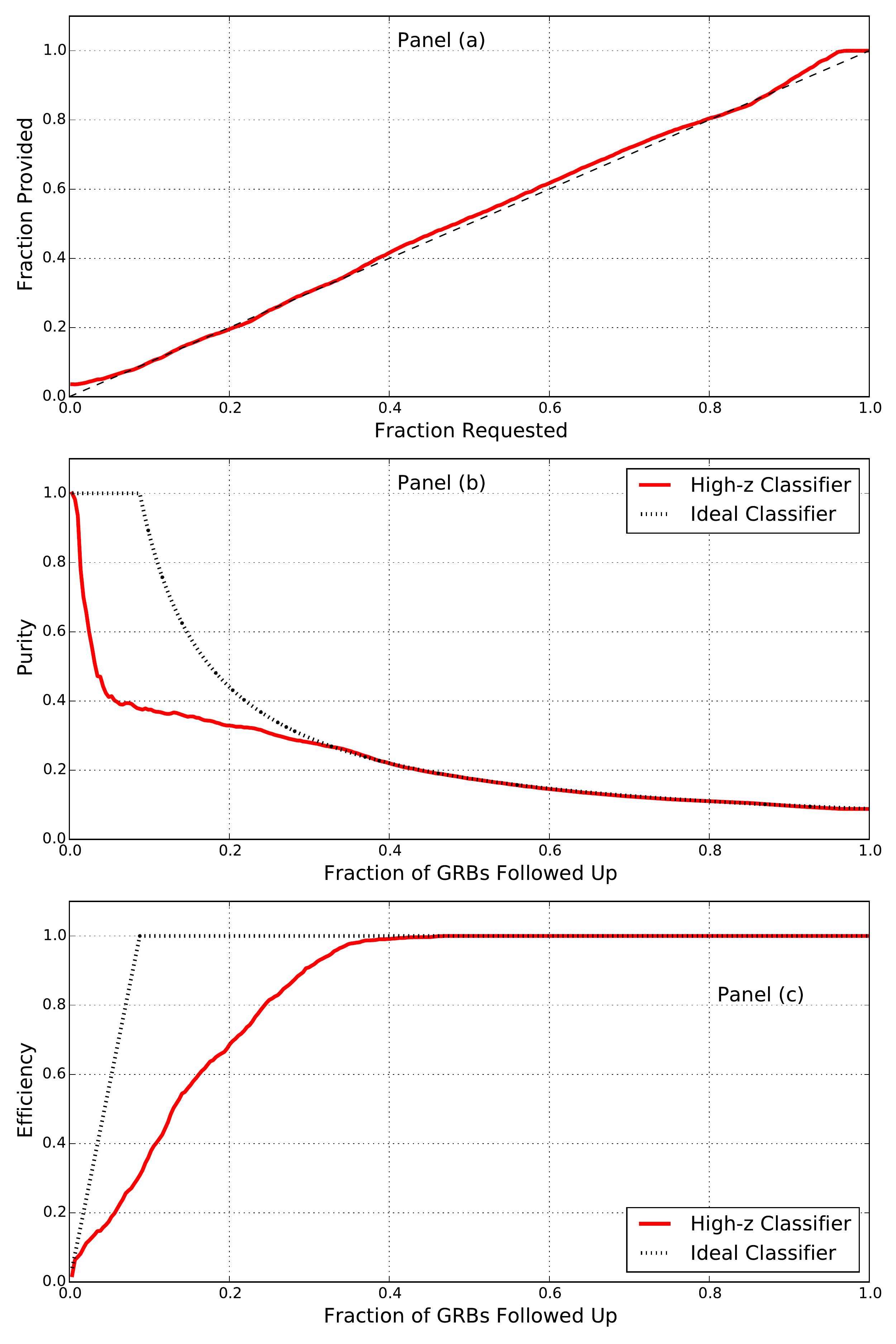}
\caption{Performance curves for {\it high-z} classifier.
The top panel (a) compares the fraction of bursts recommended for follow up
and the fraction requested from the classifier. The middle panel (b)
shows the purity of the burst sample selected for follow-up (the fraction of
bursts that were followed up that are actually at high redshift). The
bottom panel (c) shows the efficiently of the classifier (the fraction
of all high-redshift bursts that were followed up).
}\label{machine_learned_resource_allocation}
\end{figure}

We tested the above approach by simulating follow-up decisions for bursts on our training sample
using a randomized cross-validation procedure described in Section~\ref{roc_curve}.
The $Q$ scores were calculated using approximately optimized input parameters
and features found in Sections~\ref{tune_classifier} and \ref{feature_imp_classifier}.
The effectiveness of this hypothetical observing campaign as a function of the
requested follow-up fraction $F$ is shown in Fig.~\ref{machine_learned_resource_allocation}.
The top panel (a) shows that the fraction of bursts recommended for follow-up by the algorithm
($Q < F$) closely tracks the requested value $F$. The middle panel (b) shows the purity
of the follow-up sample, i.e. the number of actual high-z GRBs divided by the total
number of selected high-z candidates. Ideally, the purity would be close to 100\%
when the follow-up resources are limited (low $F$), as shown by the green line.
The bottom panel (c) shows the efficiency of selecting high-z GRBs (the fraction
of all high-z bursts that were actually observed). Again, perfect classification
performance is shown by the green line.

From Fig.~\ref{machine_learned_resource_allocation} it is clear that the classifier
can identify high-redshift bursts with a high probability, especially
when follow-up resources are very limited (low $F$). In other words, the purity
is highest when very few bursts can be followed up and therefore reliable
predictions matter most. At 1\% follow-up fraction the purity exceeds 80\%.
On the other hand, an observer with enough telescope time to follow up
40\% of all GRBs will be able to find all true high-z bursts (100\% efficiency).

%/home/tilan/Desktop/Dropbox/grb/high_z_screening/classification_morgan/10_fold/classification_analysis_10_fold_v4.py

\section{Machine-z Redshift Estimator} \label{machine_z}

The high-z classifier developed in section~\ref{highz_classifier} helps
to select the highest priority follow-up targets, but it does not provide
an actual value for the predicted redshift. In this section,
we adapt the methods from section~\ref{highz_classifier} to solve a regression problem
and develop an RF based redshift estimator that we call {\it machine-z}.

\subsection{Tuning the Regressor} \label{tune_regressor}

Since the performance of a regressor is evaluated differently from a classifier,
we performed an independent parameter search to approximately optimize
input parameters of the RF regressor. For this purpose we used the ``leave-one-out''
cross-validation method that for $N$ bursts consists of $N$ runs with $N-1$
instances used for training and one for testing. We leverage the stochastic nature
of RF training to increase the signal-to-noise ratio of the final cross-validated
performance estimate by repeating the process 10 times with different seeds.
The quality of prediction is measured using the Pearson correlation
coefficient between {\it machine-z} output and true redshift.
Table~\ref{parameters} defines the search grid for approximate parameter optimization.
In this case we found that a forrest of 100 fully developed trees (with as little as
one burst per leaf node) and $m = 5$ random features per split provides the best results.
This is different from parameters adopted in section~\ref{tune_classifier}.
The resulting correlation coefficient is 0.52.

%/home/tilan/Desktop/Dropbox/grb/high_z_screening/almost_all_analysis_feature_selection/almost_all_analysis_feature_selection_v4.py
\begin{figure*}
\includegraphics[width=0.8\textwidth]{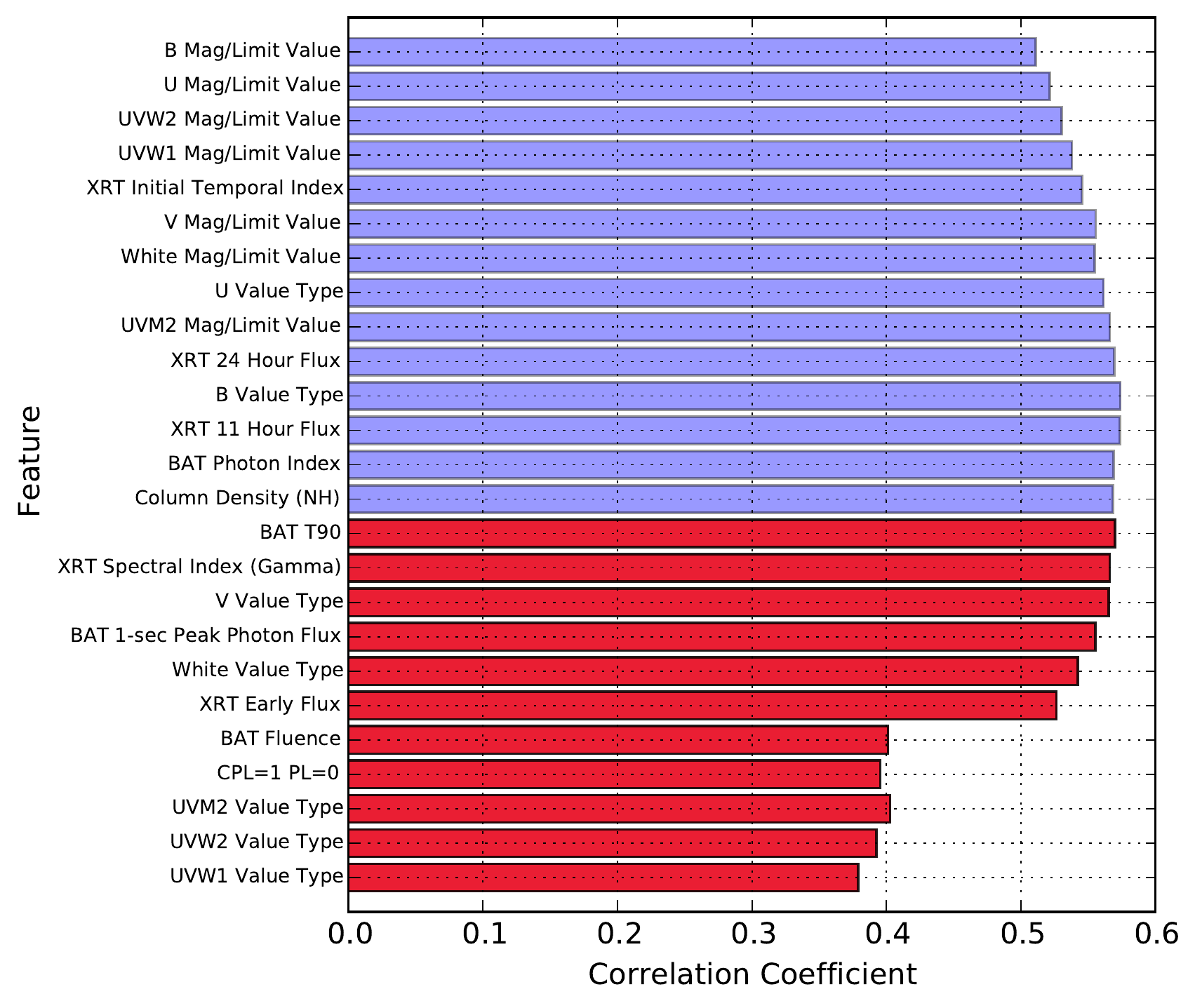}
\caption{Relative importance of {\it machine-z} regression features.
The Pearson correlation coefficient is shown as a function of the next best feature
starting from the single best feature at the bottom of the plot.
Features selected for the final {\it machine-z} estimation are shown in red.}
\label{feature_selection_regressor}
\end{figure*}

\subsection{Regression Feature Importance}\label{feature_imp_regressor}

Feature importance for {\it machine-z} is determined using a method similar
to Section~\ref{feature_imp_classifier} except for the objective function.
The area under the ROC curve is now replaced by the redshift correlation
coefficient. Until the number of selected input features reaches $m = 5$
all features are randomized during node splitting.
The relative importance of various features is shown in
Fig.~\ref{feature_selection_regressor}. The correlation
coefficient starts from a sub-optimal value for the first feature,
then increases, eventually flattens after the 11-th feature, and
then slowly decreases beyond 16-th feature. We selected the first 11 features
in this plot as input features for the {\it machine-z} estimator. Adding
features beyond 11 does not improve predictions and increases the risk
of overfitting or diluting the signal with noisy features.

\subsection{Correction for Noise and Imbalance} \label{machinez_cal}

\begin{figure}
\includegraphics[width=84mm]{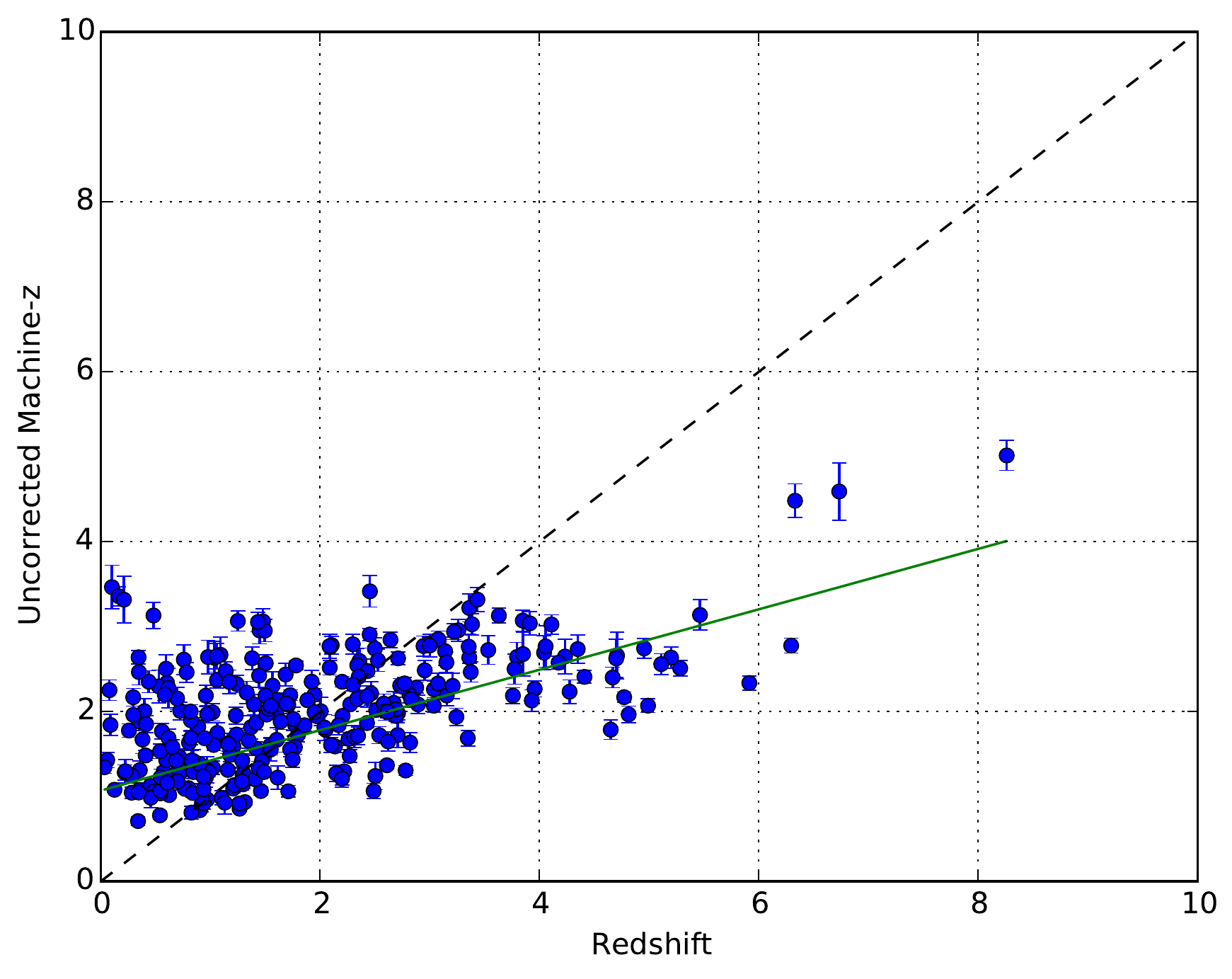}
\caption{Comparison of uncorrected {\it machine-z} predictions with true redshift.
The correlation coefficient between the two quantities is 0.57.
The best straight line fit is show in green.}
\label{almost_all_correlation}
\end{figure}

A comparison between {\it machine-z} predictions and true redshift
for GRBs in the training set is presented in Fig.~\ref{almost_all_correlation}.
While there is a good correlation between the predicted and the actual
redshift, the range of the output values is squeezed relative to the input.
This appears to be a consequence of the interaction between noisy features
and the fact that high-redshift bursts are strongly underrepresented
in training data (see Fig.~\ref{zhisto}). When high-redshift bursts are given
higher weights (e.g. by including multiple copies of the same burst in training data),
the bias observed in Fig.~\ref{almost_all_correlation} changes. A thorough
investigation of this behavior is beyond the scope of this paper and will be
presented elsewhere. For this initial release of the algorithm we introduce
a simple linear correction that shifts and stretches the range of the output
redshift values while preserving the correlation coefficient.
The final corrected redshift predictions are computed using a straight line fit
to data in Fig.~\ref{almost_all_correlation} and taking into account the cross-validation
uncertainty in {\it machine-z} output:
$z_{\rm corrected} = (z_{\rm uncorrected} - {1.07 \pm 0.05})/({0.35 \pm 0.03})$.

\begin{figure}
\includegraphics[width=84mm]{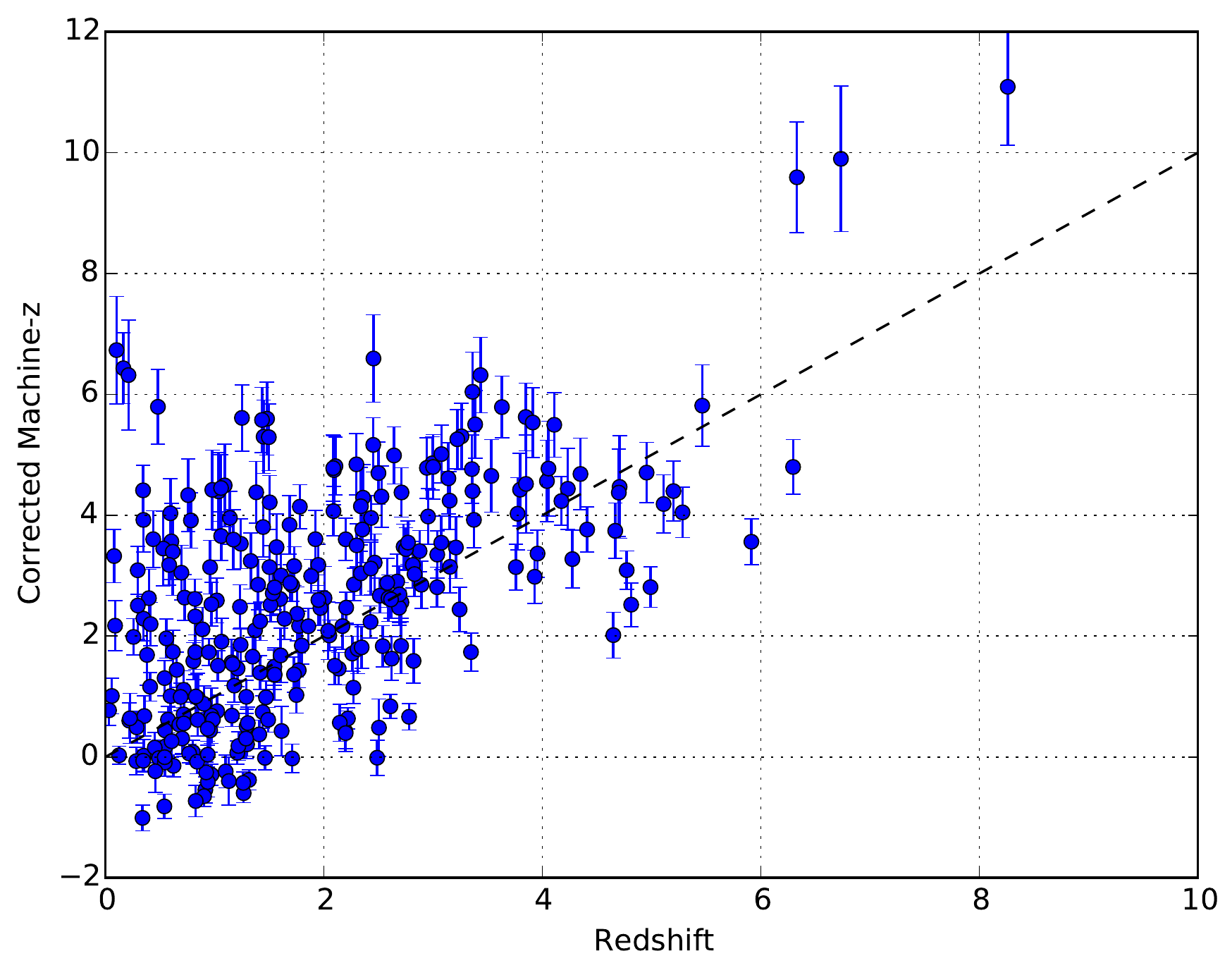}
\caption{Comparison of corrected {\it machine-z} predictions with true redshift.
The correlation coefficient between the two quantities is 0.57.}
\label{almost_all_correlation_corrected}
\end{figure}

The final corrected {\it machine-z} predictions as a function of the true redshift
are shown in Fig.~\ref{almost_all_correlation_corrected}. 
%Fig.~\ref{almost_all_2d_histogram}
%depicts the same results as a density map to show the relative number of bursts
%available for regression in different parts of the plot.
The range of the output is now similar to that of the input and the correlation 
coefficient is the same as in Figure~\ref{almost_all_correlation}.

\begin{figure}
\includegraphics[width=84mm]{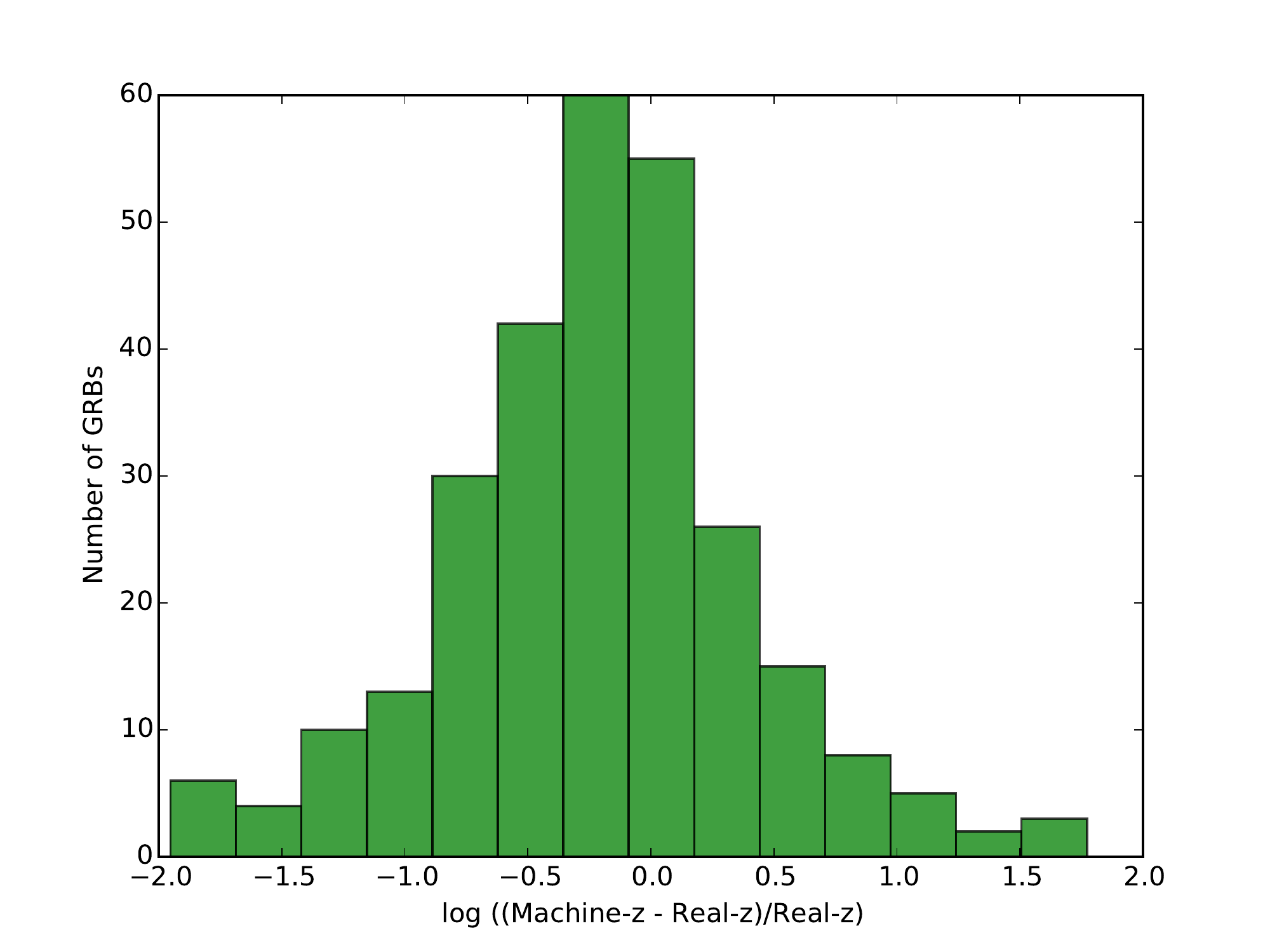}
\caption{Distribution of relative differences between
{\it machine-z} predictions and true redshifts.}\label{almost_all_histogram}
\end{figure}

There are several interesting trends to note in Fig.~\ref{almost_all_correlation_corrected}.
%and Fig.~\ref{almost_all_2d_histogram}.
First, the lower right area of the plot is not populated. This means that in most
cases {\it machine-z} does not fail to recognize a high-redshift burst. Second,
the density of bursts peaks roughly along the dashed line, so for a significant fraction
of bursts the redshift estimate is close to the true value. This can be seen more clearly
in Fig.~\ref{almost_all_histogram} showing the distribution of the relative differences
between {\it machine-z} estimates and actual redshifts.
Third, the algorithm does occasionally predict a high redshift for a low-z burst
as shown by the upper left portion of the plot. Even though following up these false
positives will tend to waste some telescope time, {\it machine-z} will rarely miss
the all important high-z bursts. An inconvenient side effect of our simple correction
for the redshift bias is that for a few GRBs the predicted redshift is negative.
This is not a problem as long as the tool is used to select high-redshift GRBs,
as the negative predictions only occurr at low redshift.

%/home/tilan/Desktop/Dropbox/grb/high_z_screening/almost_all_analysis/almost_all_analysis_v4.py

%/home/tilan/Desktop/Dropbox/grb/high_z_screening/almost_all_analysis/almost_all_analysis_v4.py

%/home/tilan/Desktop/Dropbox/grb/high_z_screening/almost_all_analysis/almost_all_analysis_heatmap_v1.py
%\begin{figure}
%\includegraphics[width=84mm]{machine_z_heatmap.pdf}
%\caption{Color coded density map of predicted versus true redshift corresponding
%to scatter plot in Figure~\ref{almost_all_correlation_corrected}.
%}\label{almost_all_2d_histogram}
%\end{figure}

%/home/tilan/Desktop/Dropbox/grb/high_z_screening/almost_all_analysis/almost_all_analysis_v4.py

\section{Discussion} \label{discussion}

\subsection{Comparison with Previous Work} \label{comparison}

%/home/tilan/Desktop/Dropbox/grb/high_z_screening/roc_curve_analysis/combine_roc_curves.py
\begin{figure}
\includegraphics[width=84mm]{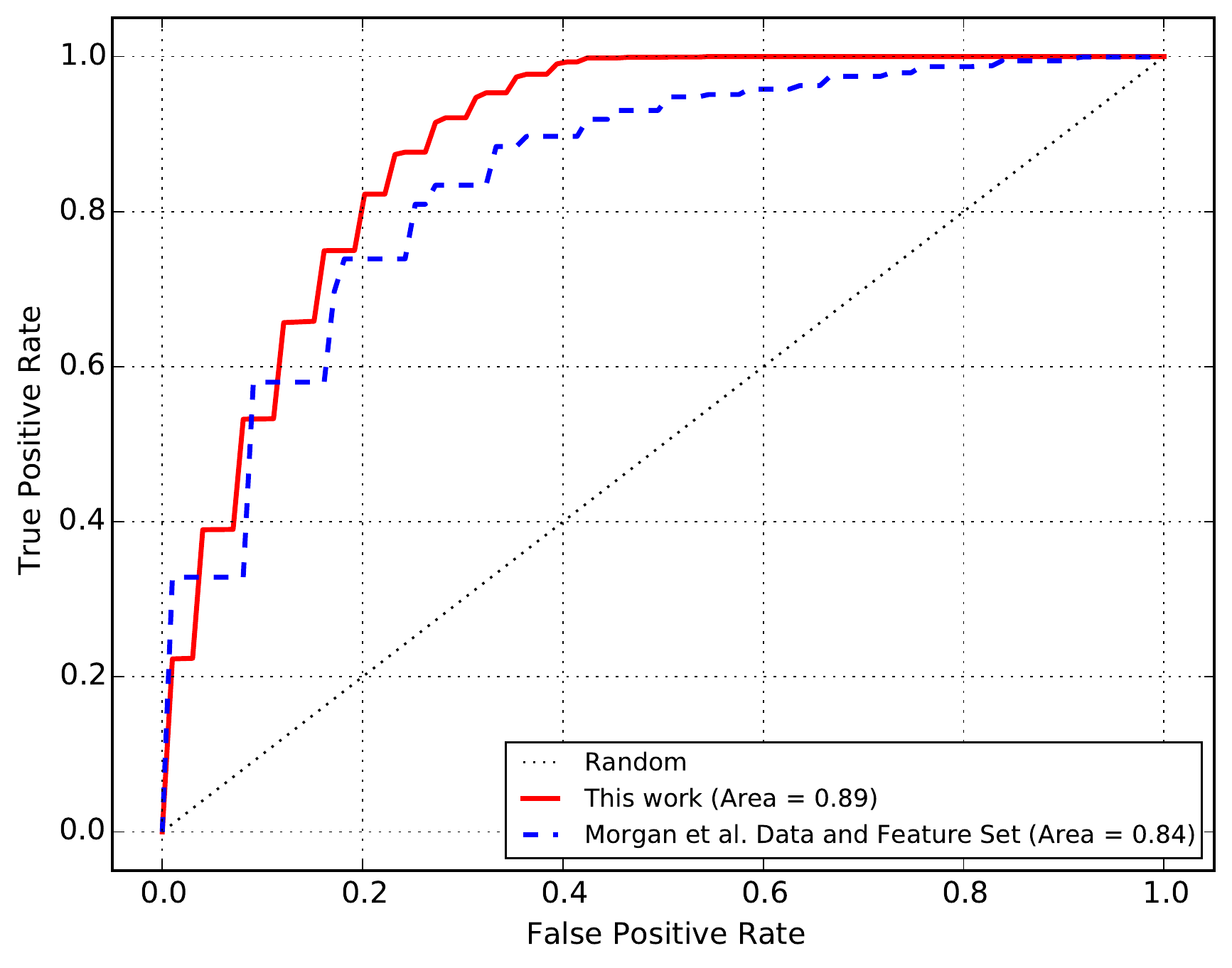}
\caption{Comparison of ROC curves for {\it high-z} classifier based on our input data
and feature set with Morgan et al. data and feature set. The area under the ROC
curve is 0.89 for this work and 0.84 for Morgan et al..
}\label{multi_roc_curve}
\end{figure}

\cite{Morgan2012} was the first to apply machine learned classification
to screen high redshift GRBs using promptly available $Swift$ data.
We compared our GRB sample and feature set with the \cite{Morgan2012}
data available in machine readable form. Small differences in the
RF implementation between those two studies have no bearing on this comparison.
The ROC curves for the two data sets are given in Fig.~\ref{multi_roc_curve}.
The ROC curve corresponding to our data set (red curve) has a slightly larger
area than that for the \cite{Morgan2012} data (blue curve).
Note that the red curve rises to 100\% recall more rapidly than the blue curve.

%/home/tilan/Desktop/Dropbox/grb/high_z_screening/classification_morgan/10_fold/classification_analysis_10_fold_v4.py
\begin{figure}
\includegraphics[width=84mm]{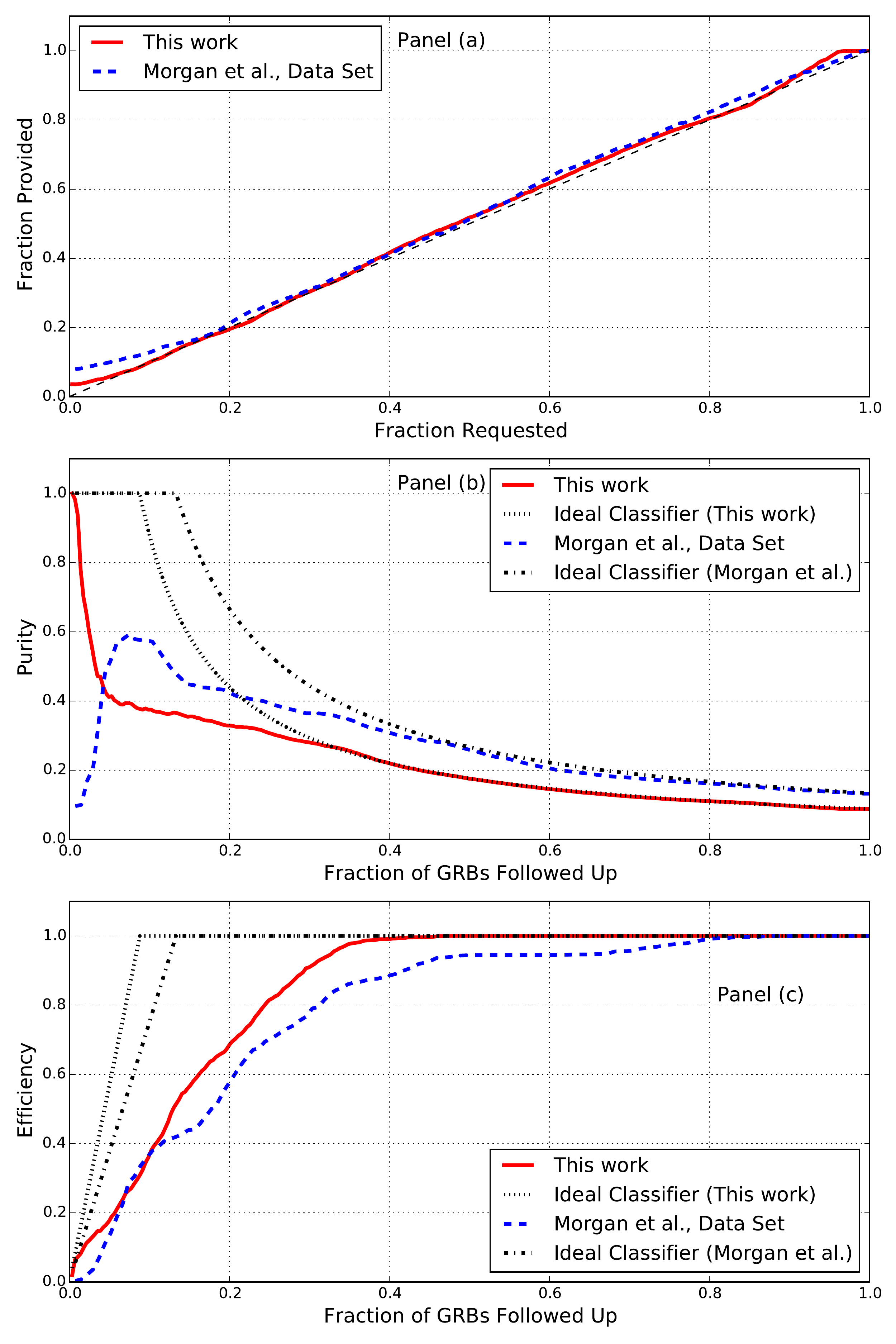}
\caption{Performance of {\it high-z} classification using our input
data and feature set compared to results based on the input data and feature set
of Morgan et al. (2012). The content of each panel is analogous
to Figure~\ref{machine_learned_resource_allocation}.}
\label{machine_learned_resource_allocation_comparison}
\end{figure}

Fig.~\ref{machine_learned_resource_allocation_comparison}
presents another performance comparison of the two data sets.
As shown by the top panel (a), there is no significant difference
in the fraction of bursts recommended for follow-up versus the requested fraction.
However, there is a significant difference in the purity of the burst sample selected for
follow-up shown in the middle panel (b). Our {\it high-z} classifier
returns samples of very high purity when the fraction of bursts that can be
observed is low. In contrast with that, the \cite{Morgan2012} data set
starts near zero purity at low followup fractions and peaks at $\approx$ 60\%
around $F = 10$\%. Furthermore, the purity curve delivered by {\it high-z}
is similar in shape to the ideal purity curve and merges with the ideal curve around
the follow-up fraction $F\, \approx$ 35\% . By comparison, the \cite{Morgan2012}
data set shows a qualitatively different shape and an undesirable very low purity
at small follow-up fractions. Finally, the bottom panel (c)
in Fig.~\ref{machine_learned_resource_allocation_comparison}
compares efficiency curves of the two data sets.
Both curves show the expected gradual rise from zero.
The {\it high-z} data set presented here reaches the curve corresponding to a perfect
classifier around follow-up fraction $F\, \approx$ 40\%. The \cite{Morgan2012}
data set does not reach the ideal curve until $\approx$ 80\%.

This difference in behavior may be explained by a different approach to integrating
XRT and UVOT measurements with non-detections. While the \cite{Morgan2012} classifier
uses 12 features, our {\it high-z} algorithm uses 8 features. Some features such as the XRT
column density are common to both data sets. However, \cite{Morgan2012}
use only one measurement from XRT and only one feature from UVOT. The latter is limited
to a yes or no flag indicating the existence of a UVOT detection in any photometric band.
Our approach, by contrast, is to use multiple features from XRT and utilize
all available information on detections and non-detections across all UVOT bands.
As one can see from Fig.~\ref{feature_selection_classifier}, both detections
and non-detections in various UVOT bands play an important role in the
{\it high-z} classifier.

\subsection{Validation} \label{validation}

\begin{table}
\centering
\caption{Validation sample: GRBs discovered in 2015 by $Swift$ with redshift measurements.
Gray rows mark candidate high-z bursts selected by both {\it high-z} classifier and {\it machine-z}
estimator.}
\label{val_cls}
\begin{tabular}{@{}lllll}
\hline \hline
GRB & Redshift & {\it Machine-z} & High-z? & Score ($Q$ value) \\
\hline \hline
\rowcolor{gray!50}151112A  & 4.1   & 5.47   $\pm$ 0.56 & True & $<$17.1 \% \\
\rowcolor{gray!50}151111A  & 3.5   & 4.81   $\pm$ 0.59 & True & $<$7.5  \% \\
151031A  & 1.17  & 3.05   $\pm$ 0.39 & True & $<$22.8 \% \\
151029A  & 1.42  & 6.1    $\pm$ 0.71 & False & $<$64.8 \% \\
\rowcolor{gray!50}151027B  & 4.06  & 5.2    $\pm$ 0.59 & True & $<$7.3  \% \\
151027A  & 0.81  & 0.32   $\pm$ 0.21 & False & $<$91.5 \% \\
151021A  & 2.33  & 3.69   $\pm$ 0.38 & True & $<$12.4 \% \\
150915A  & 1.97  & 1.38   $\pm$ 0.39 & False & $<$55.3 \% \\
150910A  & 1.36  & 2.7    $\pm$ 0.41 & False & $<$55.7 \% \\
150821A  & 0.76  & 1.37   $\pm$ 0.34 & True & $<$12.3 \% \\
150818A  & 0.28  & 0.34   $\pm$ 0.2  & False & $<$56.2 \% \\
150727A  & 0.31  & 1.13   $\pm$ 0.3  & False & $<$40.2 \% \\
150424A  & $<$3.0 & -0.28  $\pm$ 0.18 & False & $<$47.3 \% \\
150423A  & 1.39  & -0.26  $\pm$ 0.21 & True & $<$4.5  \% \\
150413A  & 3.2   & 0.75   $\pm$ 0.29 & False & $<$36.8 \% \\
150403A  & 2.06  & 2.36   $\pm$ 0.33 & False & $<$62.9 \% \\
150323A  & 0.59  & 3.81   $\pm$ 0.38 & True & $<$29.1 \% \\
150314A  & 1.76  & 2.99   $\pm$ 0.43 & False & $<$31.8 \% \\
150301B  & 1.52  & 6.06   $\pm$ 0.78 & False & $<$39.8 \% \\
150206A  & 2.09  & 2.81   $\pm$ 0.39 & False & $<$35.3 \% \\
150120A  & 0.46  & 0.81   $\pm$ 0.26 & False & $<$74.2 \% \\
150101B  & 0.09  & 0.71   $\pm$ 0.35 & False & $<$73.9 \% \\
\hline
\end{tabular}
\end{table}

The training data for our {\it high-z} classifier and {\it machine-z} estimator
is limited to GRBs discovered by $Swift$ prior to 2015 for which a
redshift measurement is available. In 2015 $Swift$ found 22 additional
bursts that have a spectroscopic redshift. We used the 2015 sample
as a validation set to investigate the effectiveness of our redshift prediction algorithms.
The results for individual bursts in the test sample are shown in Table~\ref{val_cls}.
Two out of 22 bursts (GRB 151112A and GRB 151027B) qualify as high-redshift
according to our classification in section~\ref{grb_sample} ($z > 4$).
Both algorithms flag them as having high redshift. The $Q$ scores for these
two bursts also indicate that follow-up is recommended if the requested follow-up fraction
is at least 20\% of all GRBs. In addition to these two clear cut cases, our classifier
identified 8 other bursts in the validation sample as high-z. However,
out of those 8 bursts only one (GRB 151111A) has a {\it machine-z} estimate $ z > 4$.
There are also two low-z bursts (GRB 151029A and GRB 150301B) with predicted redshifts
above 4. These outcomes are consistent with our performance estimates
and confirm the usefulness of our tools in prioritizing follow-up observations of
candidate high-redshift GRBs. We can expect that the most robust results will be obtained if both
{\it high-z} classifier and {\it machine-z} estimator predict high redshift.
In this case we would have selected three candidate high-z bursts
shown as gray rows in Table~\ref{val_cls}, two real ones and a single false positive.
Note that the false positive (GRB 151111A) is a burst with an intermediate redshift $z = 3.5$.

\subsection{Extensions to Other Data Sources} \label{extensions}

The present paper addresses redshift prediction for $Swift$ GRBs. Transfering a trained classifier
from one data set to another is very important, but typically challenging. Unfortunately, in most cases
the performance is strongly degraded even if differences between data sources are purely incidental
(e.g. slightly different energy ranges of flux measurements or different estimators of model parameters).
If the new features are qualitatively similar to the old ones, one can shift and rescale the numbers
to approximately match the distribution of each new and old feature.
This requires only a modest amount of new data and may be a productive approach for future missions
similar to $Swift$ including Space-based multi-band astronomical Variable Objects Monitor (SVOM;~\cite{Cordier2015}).
In other cases we are forced to build new training sets
and that can be time consuming.

Another possible approach would be to consider lower level data such as time-resolved spectra of prompt
GRB emission. Generic intermediate level features can be obtained for example from wavelet analysis
that captures the intrinsic structure of the data and then apply a high level classfier such as RF~\citep{Ukwatta2015wavelet}.
Those ``abstract'' features may prove more transferable from
one data set to another and may eventually facilitate early redshift prediction for GRBs
detected by future missions such as SVOM.

\section{Summary} \label{summary}

We presented a method for selecting candidate high redshift gamma-ray bursts
that can be used to prioritize follow-up observations.
The algorithm utilizes numerical and categorical features from all three instruments
onboard the $Swift$ satellite that are readily available within the first few hours after
a GRB discovery. We independently developed {\it high-z} classification
and {\it machine-z} regression tools based on algorithms and features tailored to each task.
A subset of features that provides most information to support redshift prediction
was identified in both cases. The results were validated using a small sample of recently
discovered GRBs that were not included in training data. The most robust selection
of high redshift bursts is achieved by combining the classification and regression
output to make the final prediction.

\section*{Acknowledgments}

This work was funded by the US Department of Energy. We acknowledge
support from the Laboratory Directed Research and Development program at
Los Alamos National Laboratory. We also thank the anonymous referee
for comments that significantly improved the paper.

\bibliographystyle{plainnat}
\bibliography{my_references}

\bsp

\label{lastpage}

\end{document}